\documentclass[conference]{IEEEtran}
\IEEEoverridecommandlockouts
\usepackage{amsmath,amssymb,amsfonts}
\usepackage{algorithmic}
\usepackage{graphicx}
\usepackage{textcomp}
\usepackage{xcolor}
\usepackage{comment}
\def\BibTeX{{\rm B\kern-.05em{\sc i\kern-.025em b}\kern-.08em
    T\kern-.1667em\lower.7ex\hbox{E}\kern-.125emX}}

\usepackage{tabularx}
\usepackage{caption}

\usepackage{textcmds}

\usepackage{{booktabs}}

\usepackage{tcolorbox}

\usepackage{enumerate}
\usepackage[numbers,sort&compress]{natbib}

\usepackage{subcaption}
\newsavebox{\largestimage}

\usepackage{soul}
\usepackage{tikz}
\usetikzlibrary{calc}
\usepackage{hyperref}
\usepackage{cleveref}

\usepackage[normalem]{ulem}

\makeatletter
\newif\if@anonymize

\@anonymizefalse  %

\if@anonymize
  \newcommand{\highlight@DoHighlight}{
    \fill [outer sep = -15pt, inner sep = 0pt, color=black]
          ($(begin highlight)+(0,8pt)$) rectangle ($(end highlight)+(0,-3pt)$) ;
  }

  \newcommand{\highlight@BeginHighlight}{
    \coordinate (begin highlight) at (0,0) ;
  }

  \newcommand{\highlight@EndHighlight}{
    \coordinate (end highlight) at (0,0) ;
  }

  \newdimen\highlight@previous
  \newdimen\highlight@current
  \newlength{\item@width}

  \DeclareRobustCommand*\anonymize{%
    \SOUL@setup
    \def\SOUL@preamble{%
      \begin{tikzpicture}[overlay, remember picture]
        \highlight@BeginHighlight
        \highlight@EndHighlight
      \end{tikzpicture}%
    }%
    \def\SOUL@postamble{%
      \begin{tikzpicture}[overlay, remember picture]
        \highlight@EndHighlight
        \highlight@DoHighlight
      \end{tikzpicture}%
    }%
    \def\SOUL@everyhyphen{%
      \discretionary{%
        \SOUL@setkern\SOUL@hyphkern
        \SOUL@sethyphenchar
        \tikz[overlay, remember picture] \highlight@EndHighlight ;%
      }{%
      }{%
        \SOUL@setkern\SOUL@charkern
      }%
    }%
    \def\SOUL@everyexhyphen##1{%
      \SOUL@setkern\SOUL@hyphkern
      \settowidth{\item@width}{##1}%
      \makebox[\item@width]{}%
      \discretionary{%
        \tikz[overlay, remember picture] \highlight@EndHighlight ;%
      }{%
      }{%
        \SOUL@setkern\SOUL@charkern
      }%
    }%
    \def\SOUL@everysyllable{%
      \begin{tikzpicture}[overlay, remember picture]
        \path let \p0 = (begin highlight), \p1 = (0,0) in \pgfextra
          \global\highlight@previous=\y0
          \global\highlight@current =\y1
        \endpgfextra (0,0) ;
        \ifdim\highlight@current < \highlight@previous
          \highlight@DoHighlight
          \highlight@BeginHighlight
        \fi
      \end{tikzpicture}%
      \settowidth{\item@width}{\the\SOUL@syllable}%
      \makebox[\item@width]{}%
      \tikz[overlay, remember picture] \highlight@EndHighlight ;%
    }%
    \SOUL@
  }
\else
  \newcommand{\anonymize}[1]{#1}
\fi
\makeatother

\begin{document}

\title{Predicting Defective Visual Code Changes in a Multi-Language AAA Video Game Project
}

\author{\IEEEauthorblockN{\anonymize{Kalvin Eng}}
\IEEEauthorblockA{
\anonymize{\textit{Quality, Verification \& Standards}} \\
\anonymize{\textit{Electronic Arts}}\\
\anonymize{Edmonton, Canada} \\
\anonymize{kalvin.eng@\{ualberta.ca, ea.com\}}}
\and
\IEEEauthorblockN{\anonymize{Abram Hindle}}
\IEEEauthorblockA{
\anonymize{\textit{Department of Computing Science}} \\
\anonymize{\textit{University of Alberta}}\\
\anonymize{Edmonton, Canada} \\
\anonymize{abram.hindle@ualberta.ca}}
\and
\IEEEauthorblockN{\anonymize{Alexander Senchenko}}
\IEEEauthorblockA{
\anonymize{\textit{Quality, Verification \& Standards}} \\
\anonymize{\textit{Electronic Arts}}\\
\anonymize{Vancouver, Canada} \\
\anonymize{asenchenko@ea.com}}
}

\maketitle

\begin{abstract}
Video game development increasingly relies on using visual programming languages as the primary way to build video game features. The aim of using visual programming is to move game logic into the hands of game designers, who may not be as well versed in textual coding. In this paper, we empirically observe that there are more defect-inducing commits containing visual code than textual code in a AAA video game project codebase. This indicates that the existing textual code Just-in-Time (JIT) defect prediction models under evaluation by \emph{Electronic Arts} (EA) may be ineffective as they do not account for changes in visual code. Thus, we focus our research on constructing visual code defect prediction models that encompass visual code metrics and evaluate the models against defect prediction models that use language agnostic features, and textual code metrics. We test our models using features extracted from the historical codebase of a AAA video game project, as well as the historical codebases of 70 open source projects that use textual and visual code. We find that defect prediction models have better performance overall in terms of the area under the ROC curve (AUC), and Mathews Correlation Coefficient (MCC) when incorporating visual code features for projects that contain more commits with visual code than textual code.

\end{abstract}


\section{Introduction}
\begin{figure}[tbp]
    \centering
    \includegraphics[width=0.45\textwidth]{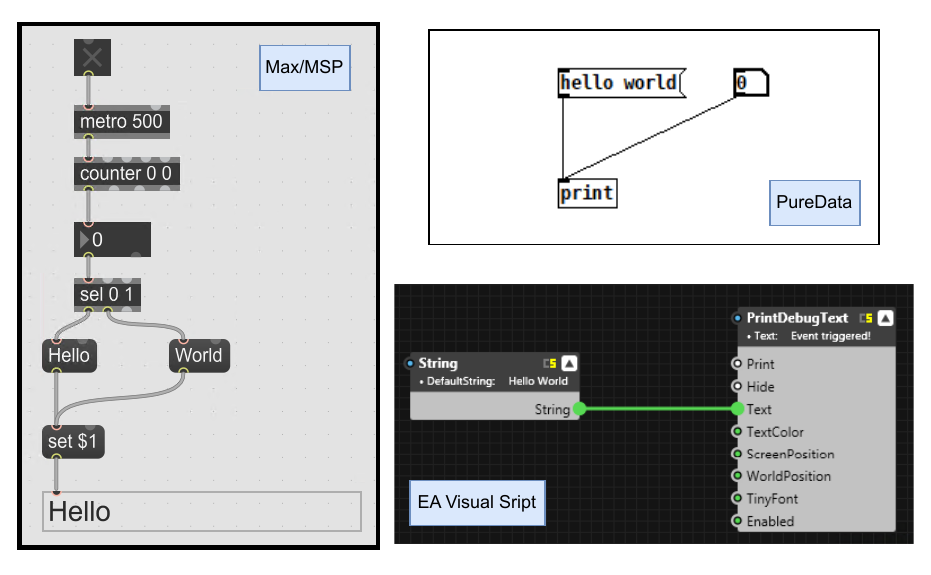}
    \caption{Visual code examples: Max/MSP, Pure Data, and EA Visual Script.}
    \label{fig:vc-example}
\end{figure}

Visual code, also known as block-based code, low code, no-code, or visual scripts, is a type of software development practice that aims to simplify source code for non-traditional programmers by representing code visually using drag and drop nodes connected via edges, instead of using only text. Non-traditional programmers are end-users who are not well-versed in writing in line-by-line textual code~\cite{ko2011state}. Visual code, illustrated in \Cref{fig:vc-example}, attempts to simplify the software development process for non-traditional \textit{end-user programmers} by making dataflow clear and avoiding some classes of syntax errors. Traditional \textit{professional programmers}, in comparison, mainly write software with textual code, and have adequate tool support. 

At \emph{EA} (Electronic Arts), video game development teams have shifted towards using visual code as the primary way to build video game features, such as triggers on maps, or game rules, so that all parts of a team can be involved in the software development process. A video game development team is diverse and includes managers, developers, designers, artists, and testers. With many stakeholders, the goal of visual code is to allow simple changes to code such as using different variables or equations, without waiting for or consulting with traditional \textit{professional programmers}. Popular implementations of visual code occur in game development include Unreal Engine Blueprints~\cite{unreal-blueprints}, and Unity Visual Scripting~\cite{unity-vs}.

Due to the introduction of non-traditional \textit{end-user programmers} using visual code in video game projects, we wish to investigate how to predict defects in visual code. Using the development history of a AAA video game project (a big budget game from a large studio) along with 70 open source visual programming projects to build defect prediction models, we answer the following research questions:
\begin{enumerate}[{RQ1.}]\bfseries
  \item Do defect-inducing commits with visual code occur more frequently than defect-inducing commits with textual code in the projects studied?
  \item Do visual code features significantly improve the performance of defect prediction models for textual/visual code projects?
  \item Do the types of files in commits or choice of learners in projects significantly affect the outcome of defect prediction models?
\end{enumerate}

The goal of this research is to investigate whether contextual features that describe visual code improve defect prediction in projects that use textual and visual code. Visual code metrics, to the best of our knowledge, have never been used for predicting the likelihood of defects.

\section{Background}
Defect prediction models are classification models that estimate the likelihood of a change in code being defective~\cite{xai4sebook}. Tantithamthavor \textit{et al.}~\cite{xai4sebook} explain that defect prediction models have been used to predict the likelihood of defects at different granularities including: packages~\cite{kamei2010revisiting}, components~\cite{thongtanunam2016revisiting}, modules~\cite{kamei2007effects}, files~\cite{kamei2010revisiting, mende2010effort}, and methods~\cite{hata2012bug, pascarella2020performance}. Defect prediction models have also been used at version control commit-level~\cite{pascarella2019fine, nayrolles2018clever, rosen2015commit}. The use of defect prediction models at the time of code commits can be referred to as \textit{Just-In-Time} (JIT) defect prediction and can be leveraged immediately once a change is committed to a repository~\cite{kamei2012large}. JIT defect prediction is often used to better understand where to deploy testing efforts~\cite{kamei2012large}. EA uses JIT defect prediction models to perform focused code reviews and testing efforts on specific changes of code at a commit level~\cite{senchenko2022supernova}. The focus of this research is to improve current EA JIT defect prediction models by including features that involve visual code metrics in addition to process metrics, and textual code metrics.

\textit{Process metrics} capture information about changes during the software development process and aims to describe the relationship between changes and software quality~\cite{xai4sebook}. It is language agnostic and can be applied uniformly to software written in different languages~\cite{majumder2022revisiting}. Examples of process metrics may involve: number of revisions for a file, number of developers for a file, number of modified lines, and number of directories~\cite{majumder2022revisiting, madeyski2015process}. According to \citet{majumder2022revisiting}, most studies investigating the use of process metrics show that models that only use process metrics can outperform those that combine process metrics and code metrics. However, \citet{kamei2010revisiting} finds that combining process metrics and code metrics improves defect prediction performance.

\textit{Code metrics}, also referred to as product metrics, capture information about code and aims to describe the relationship between code properties and software quality~\cite{xai4sebook}. Examples of code metrics can include: lines of code in a file, cyclomatic complexity of code in a file, and number of methods in a file~\cite{majumder2022revisiting, gyimothy2005empirical, madeyski2015process}.
In this paper, we separate code metrics into 2 categories: \textit{textual} which refers to the metrics described in this paragraph that are derived from textual code and \textit{visual} which refers to metrics derived from visual code.

We define visual code metrics as metrics that capture information about visual code to describe the relationship between visual code properties and software quality. Visual code metrics have been investigated by \citet{kumar2016source} who propose metrics derived from the operators and operands of visual source code defined by the IEC 61131-3 Programmable Logic Controller programming languages standards. Using the same programming languages, visual code metrics have also been investigated by \citet{fischer2021measuring} who propose metrics that measure complexity based on size, data structure, control flow, information flow and lexical structure derived in the textual and visual source code. The goal of this investigation is to derive a suite of visual code metrics suitable for defect prediction in visual code changes. 

\subsection{Choices in JIT Defect Prediction}
\label{subsec:defect-pred-challenge}
Defect prediction is all about choices that can impact the outcomes of defect prediction models. Since we are concerned about \emph{Just In Time} (JIT) defect prediction, we outline the process summarized in Zhao \textit{et al.}~\cite{zhao2023systematic} that includes choices in (1) data acquisition, (2) data preparation, (3) model building, and (4) model evaluation.

\textbf{(1) Data acquisition} concerns the sources of data for a defect prediction model which includes retrieving software change history from version control systems and classifying the software changes as defect-inducing or clean~\cite{zhao2023systematic}. To mark a software change as defect-inducing, defect-fixing changes can be identified from issue tracking systems or commit messages. Defect-fixing changes can be linked to defect-inducing changes using variants of the SZZ algorithm~\cite{sliwerski2005changes, szz-vc}.
The choice of how defect-inducing changes are found can affect the outcomes of a defect prediction model~\cite{quach2021evaluating, fan2019impact}.

\textbf{(2) Data preparation} encompasses feature acquisition and processing, that is acquiring the sets of metrics from the change history data that will be used for the defect prediction model~\cite{zhao2023systematic}. Features can be acquired from numerous sources including: software changes, commit messages, issue tracking systems, static analysis, and by automatically learning using algorithms like deep learning~\cite{zhao2023systematic}. Furthermore, features can be extracted at different levels of granularity depending on what granularity of defect prediction a model will be (e.g.\ file-level vs commit-level). As well, these extracted features need to be preprocessed including dealing with skew, collinearity and multicollinearity, and class imbalance~\cite{zhao2023systematic}. Hence, there are many choices to be made about which features to include in a model, and how to preprocess data for building the model.

\textbf{(3) Model building} includes deciding on the granularity of defect prediction, choice of learner(s), %
and what data to use~\cite{zhao2023systematic}. The granularity of a defect prediction model is a decision that impacts how interpretable a prediction is. A coarse granularity level would mean prediction at a commit-level, while a fine granularity level would be prediction at a file-level. \citet{wan2018perceptions} mentions the conclusion of \cite{kamei2016defect} that finds the \qq{practical value of prediction decreases as the granularity level increases}, i.e., opting for a coarser granularity may be more practical as it reduces the necessity of reviewing a substantial number of files. 

The choice of a learner can impact defect prediction performance~\cite{ghotra2015revisiting}, thus multiple learners should be tried.
There are many popular choices of learners including: Logistic Regression, Tree-based models (e.g.\ Random Forest, and C4.5 Decision Tree) and ensemble models (e.g.\ Random Forest, and XGBoost) which affect the performance of defect prediction results~\cite{zhao2023systematic}.

The choice of data used to build the model is also important which includes many factors such as accounting for concept drift, verification latency, defect types, and imbalanced properties of data~\cite{zhao2023systematic}. Concept drift refers to how regularities in the software change data gradually change or shift~\cite{zhao2023systematic}. To address concept drift, it is suggested that different slices of data should be used for training~\cite{mcintosh2018fix}.

Another issue to consider is verification latency which refers to the \qq{lag time between when a defect-inducing change is committed to the [version control system] and identified as such}~\cite{zhao2023systematic}. An oversampling technique is suggested to address verification latency~\cite{cabral2019class}. Defect types should also be considered since there are defect-fixing changes for \qq{extrinsic} defects that occur externally to the code. If extrinsic defects are used to find defect-inducing changes, it can negatively impact model performance as the changes have nothing to do with the fix~\cite{rodriguez2020watch}. 

Finally, class imbalance, between classes such as defect-inducing and clean, must also be considered. For example, \citet{jiang2013personalized} develop a local defect prediction model for defect prediction as they theorize that different developers exhibit different software change patterns. In a local model, a model is created for a subset of data (e.g.\ a local model for developers means that each developer will use a different model). In contrast, a global model does not account for subsets of data. It should be noted that local models might under-perform global models in defect prediction~\cite{yang2019local}.

\textbf{(4) Model evaluation} refers to how a model's performance is measured and encompasses the validation method used, the choice of evaluation metrics, and whether or not to evaluate feature importance. Validation methods can include cross-validation, however it is suggested that the defect prediction is time-sensitive and therefore data should be split on time instead~\cite{bangash2020time}. Choices of popular model evaluation metrics include: accuracy, precision, recall, F-measure, AUC, and MCC~\cite{zhao2023systematic, yao2020assessing}. Different measures should be considered depending on the needs of a defect prediction model and the importance of measuring misclassifications~\cite{hall2011systematic}. Feature importance is another dimension that can be explored, this refers to how important a feature is to explain the predicted outcome of a model. Techniques such as LIME~\cite{ribeiro2016should}, SHAP~\cite{lundberg2017unified}, and PyExplainer~\cite{pornprasit2021pyexplainer} can be used for feature explanations~\cite{xai4sebook}.

\subsection{Investigating Visual Code Defect Prediction}
We develop a commit-level visual code defect prediction model, that supports the ongoing EA evaluation of quality assurance which already uses commit-level textual code Just-in-Time (JIT) defect prediction. To encourage replication of our work, we implement a number of visual code defect prediction models using open source projects. Our choices for our visual code defect prediction models are based on the background in \Cref{subsec:defect-pred-challenge}. Our \textit{data collection} choices are introduced in \Cref{datasets}. Our category of features is explained in \Cref{data-prep}. Our method for preprocessing the data and training the defect predictors is explained in \Cref{build-model} and evaluated in \Cref{analyze-model} answering \textbf{RQ1}, \textbf{RQ2}, and \textbf{RQ3}.

\section{Datasets}
\label{datasets}
To build a defect prediction model for evaluation, we use 2 datasets: a AAA video game project, and an open source projects dataset. Our AAA video game project is chosen because it is a project that was primarily developed using visual code with 38\% of commits consisting of visual code. 
Our open source projects dataset consists of 70 projects with each project containing a combination of Max/MSP visual code and textual code. The open source projects dataset is used to motivate replication of our work, and to demonstrate generality and feasibility of using visual code features for defect prediction in projects with visual code and textual code. 

\subsection{AAA Video Game}
\begin{figure}[tbp]
    \centering
    \includegraphics[width=0.45\textwidth]{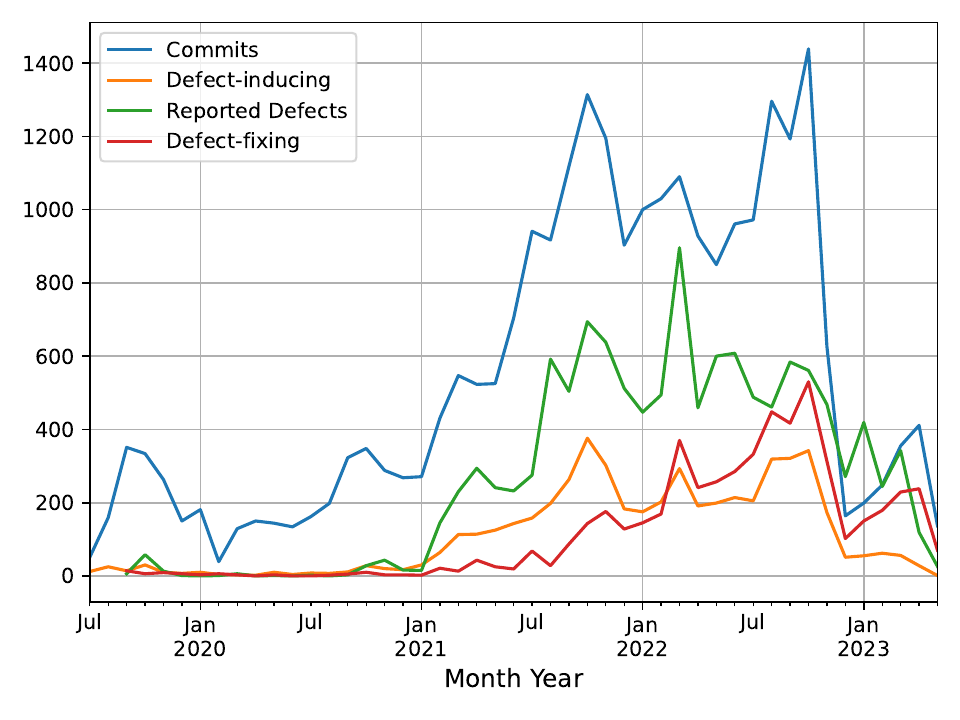}
    \caption{AAA Video Game project activity over time.}
    \label{fig:defects-month}
\end{figure}

In the AAA video game, there are 26,326 commits as of May 14, 2023, where 5,296 are defect-fixing commits, and 5,184 are defect-inducing commits. Defect-inducing commits were found using textual SZZ~\cite{kim2006automatic} for any textual code, while SZZ-VC (max change-depth)~\cite{szz-vc} (an SZZ method for finding defect-inducing changes in visual code) is used for any visual code. To find the defect-fixing changes to identify defect-inducing changes, we consider in the AAA video game project commits that are linked to a fix in the issue tracker.
We plot this project activity over time in \Cref{fig:defects-month}.

This game has been in development since July 2019 and is still actively maintained. The project is forked from a different project that has been in development since March 2014 and uses 399 of its commits.
We retrieve our code change commits from Perforce, while issue reports are extracted from Jira. It should be noted that our Jira issue tracker has been designed to work under a quality assurance workflow where defect-fixing commits are recorded in the issue as well as the version of when a defect was first found.

\subsection{Open Source Max/MSP Visual Code Projects}
Max/MSP visual code is a popular visual music programming language that is widely used by computer music programmers to realize their music compositions~\cite{burlet2015empirical}. Burlet \textit{et al.}~\cite{burlet2015empirical} describe Max/MSP as a language that allow users to programmatically arrange rectangular objects (nodes) on the screen and connect them with lines (edges) called patch cords to generate sound and respond to human-computer interaction devices. We choose Max/MSP projects because it uses nodes and edges similar to the visual code of the AAA video game project.

To find projects that contain visual code and textual code, we use the U version of the World of Code (WoC)~\cite{ma2019world} to discover 7,033 initial projects containing a commit in its history with at least one Max/MSP file with the file extension \textit{.maxpat} or \textit{.maxhelp} and includes \textit{\qq{patcher}} in its source code. It should be noted that the projects chosen are the \qq{most central} repository to represent a group of repositories found with the Louvain community detection algorithm in the WoC and does not represent every project in existence on GitHub~\cite{Mockus_2020_P}.

We set criteria for projects that we would study:
\begin{itemize}
\item They must have at least 200 commits on their main branch (as suggested by Shrikanth \textit{et al.}~\cite{nc2023assessing}) after parsing with PyDriller~\cite{PyDriller};
\item Projects must be textual and visual with at least 1 visual commit and 1 textual commit;
\item Projects must have at least 1 defect-fixing commit (by matching commit messages using \textit{perfective} keywords from \citet{rosen2015commit}), 1 visual code defect-inducing commit using (SZZ-VC (max change depth)~\cite{szz-vc}), and 1 textual code defect-inducing commit (using textual SZZ~\cite{kim2006automatic}).
\end{itemize}

As a result, this paper uses \textbf{70} open source projects consisting of \textbf{64,246} commits (\textbf{8,189} defect-fixing commits and \textbf{13,002} defect-inducing commits) for evaluating visual code defect prediction.

The \textbf{70} projects chosen contain a wide variety of commits with textual code including C, C++, C\#, Java, JavaScript, Lua, Objective-C, PHP, Python, Ruby, Swift, and TypeScript along with Max/MSP visual code. This mix of textual and visual code in commits is similar to the AAA video game. The mined open source project data is available for download in our replication package~\cite{reppackage}.

\section{Defect Prediction Features}
\label{data-prep}
Our study aims to see how visual code metrics affect commit-level defect prediction, hence we choose 3 categories of metrics for features: process metrics, textual code metrics, and visual code metrics. The process metrics are similar to the ones used by~\citet{kamei2012large} and Madeyski \textit{et al.}~\cite{madeyski2015process}.

\subsection{Process Metric Features}
The process metric features are derived from the version control system. We derive our features from valid textual code files and valid visual code files. This derivation of features by file type is similar to the datasets of \citet{ni2022just} and McIntosh \textit{et al.}~\cite{mcintosh2018fix} which appear to derive metrics for only valid file types. We present each of the process metrics below along with their intuition.

\subsubsection*{\uline{Total Modified File Sizes}} By measuring the total size of the code, the larger the file sizes, the greater the potential for defects.
\subsubsection*{\uline{Average Modified File Sizes}} By measuring the average size of the code, the intuition is that with larger file sizes, the greater the potential for defects.
\subsubsection*{\uline{Number of Unique Modified Directories}~\cite{kamei2012large}} The more directories that a change touches, the greater the potential for defects.
\subsubsection*{\uline{Average Depth of Directories}~\cite{kamei2012large}} The deeper the directories in a change, the greater the potential for defects as the change becomes more complex (deeper directories indicate more submodules).
\subsubsection*{\uline{Number of Files Modified}~\cite{kamei2012large}} The more files that are modified, the more complex the change and the greater the potential for defects.
\subsubsection*{\uline{Average Elapsed Time Since Last Commit of Modified File}~\cite{kamei2012large}} The less time between commits, the greater the potential for defects.
\subsubsection*{\uline{Average Number of Revisions Per File}~\cite{madeyski2015process}} The more revisions, the greater the potential for defects, because more changes mean that a file is more complex.
\subsubsection*{\uline{Number of Developers}~\cite{kamei2012large}} The more developers that have touched a commit, the greater the potential for defects, as each developer can have different ideas about the code.
\subsubsection*{\uline{Number of Unique Changes}~\cite{kamei2012large}} The more changes per commit, the more information a developer needs to keep track of, meaning more potential for defects.
\subsubsection*{\uline{Developer Experience}~\cite{kamei2012large}} By measuring the number of changes that a developer has made since the beginning of time, more changes indicate more experience and, therefore, a lower likelihood of a defect.
\subsubsection*{\uline{Defect-Fixing}~\cite{kamei2012large}} If the change is defect-fixing, then it is unlikely to be a defect.

\subsection{Textual Code Metric Features}
These metrics provide context about changes to visual code. We present each of the textual code metrics below along with their intuition.

\subsubsection*{\uline{Total Lines of Code Added}~\cite{kamei2012large}} The more lines that are added, the greater the potential for a defect.
\subsubsection*{\uline{Total Lines of Code Deleted}~\cite{kamei2012large}} The more lines that are deleted, the more potential there is for a defect.
\subsubsection*{\uline{Total Lines of Code Before Change}~\cite{kamei2012large}} The larger a textual code file, the greater the potential for a defect.
\subsubsection*{\uline{Code Entropy}~\cite{kamei2012large}} To measure the amount of change across files, we use the same modified lines formula in~\cite{kamei2012large}. The higher the number, the larger the change distributed across many files, implying more potential for a defect.

\subsection{Visual Code Metric Features}
The metrics presented below are used to provide context about visual code changes in a commit. The visual code metrics selected are measures that can be used across different visual programming languages, and is meant to represent changes about visual code. We use nodes in-place of lines as visual code is a node and edge based programming language. 

\subsubsection*{\uline{Total Number of Nodes Added}} The more nodes that are added, the more potential there is for a defect.
\subsubsection*{\uline{Total Number of Nodes Modified}} The more modified nodes, the greater the potential for a defect.
\subsubsection*{\uline{Total Number of Nodes Deleted}} The more deleted nodes, the greater the potential for a defect.
\subsubsection*{\uline{Number of Nodes Before Change}} The larger a visual code program, the greater the potential for a defect.
\subsubsection*{\uline{Node Modification Entropy}} Similar to the \textit{Code Entropy} textual code metric, we measure the amount of changes across files by using the code entropy formula in~\cite{kamei2012large} with nodes instead of lines. Larger values indicate a larger distribution of node changes across many files.

\section{Building and Evaluating Defect Predictors}
\label{build-model}

To build defect predictors for the AAA video game, we choose the classic 80/20 non-random split of data where 80\% of data is used for training and 20\% is used for testing. For the 70 projects of the open source projects dataset, we build predictors for each project and also choose an 80/20 split. Since our split is non-random, it is also time-aware, ensuring that the testing data will never precede any of the training data.

\subsection{Mitigating Collinearity and Multicollinearity}
To address collinearity and multicollinearity among features, we use AutoSpearman~\cite{jiarpakdee2018autospearman} with the correlation threshold of $0.7$ and Variance Inflation Factor of $5$ to automatically select our features in the training data.

In the AAA video game project, we find that among the process metrics only the number of developers and whether or not if a commit is defect-fixing are non-correlated. Among the textual code metrics, the number of lines added and the code entropy are non-correlated. For the visual code metrics, the number of nodes added and the node entropy are non-correlated. We use the groups of these features to build our defect predictors for the AAA video game.

In the open source projects dataset, we also automatically select features for each project with AutoSpearman and ensure that there is at least 1 feature in each of the 3 categories (process metrics, textual code metrics, and visual code metrics) outlined in \Cref{data-prep}.

\subsection{Addressing Class Imbalance}
To address the class imbalance in the training data of the AAA video game project and open source projects dataset, we apply SMOTE~\cite{chawla2002smote} to the training data after mitigating collinearity and multicollinearity among features. The optimal choice for addressing class imbalance is undecided in prior literature, but SMOTE is a widely used technique~\cite{zhao2023systematic}.

\subsection{Learners}
\label{learners}
We choose a wide variety of learners including ones with interpretability:  \emph{Logistic Regression} (LR - Linear Statistical technique), \emph{C5.0} (DT - Decision Tree), and ones that have been demonstrated to be higher performing learners: \emph{Random Forests} (RF - Tree), \emph{XGBoost} (XGB - Tree), \emph{Gradient Boosting Method} (GBM - Tree), \emph{Multi-layer Perceptron} (NN - Neural Network). For each of the learners, we use the default parameters of learners with the \texttt{scikit-learn} version 1.2.2 library~\cite{scikit-learn} and xgboost version 1.7.6 library~\cite{Chen:2016:XST:2939672.2939785}.

\subsection{Evaluating Learners and Features for Defect Prediction}
\label{evalaute-model}
We evaluate the effect of defect prediction features by using 6 learners outlined in \Cref{build-model} on 4 different combinations of features: (1) \emph{process metrics only} (Base), (2) \emph{process metrics with textual code metrics} (Textual), (3) \emph{process metrics with visual code metrics} (Visual), and (4) \emph{process metrics with textual code metrics and visual code metrics} (Combined).

Since we train the defect prediction models using the 6 learners for each feature combination, this produces 24 defect prediction models per project. In total there are 24 defect prediction models trained for the AAA video game project, and 1,680 defect prediction models for the open source projects dataset.

We evaluate the performance of each defect prediction model with 2 lenses on performance: \emph{area under the  ROC Curve} (AUC) and \emph{Matthews correlation coefficient} (MCC). AUC measures the area under the plot of the true positive rate vs the true negative rate and ranges between 0 and 1, where a 0 indicates that a model is perfectly incorrect, 0.5 indicates that a model is randomly guessing, while a 1 indicates that a model is perfectly correct. MCC measures the correlation between the predicted values and actual values and ranges between -1 and +1, where -1 indicates no agreement, 0 indicates no correlation at all, and +1 indicates perfect agreement. AUC is used for measuring model discriminatory power, while MCC is used for measuring model correctness.

To rank the defect prediction models, we use the \emph{Non-Parametric ScottKnott ESD} (NPSK) test. The NPSK test is a multiple comparison approach that uses hierarchical clustering to partition the set of medians of model performance into statistically distinct ranked groups that are not significantly different or have an insignificant effect size~\cite{tantithamthavorn2017mvt, tantithamthavorn2018optimization}. We use the \textit{ScottKnottESD} R package~\cite{npsk-r} to perform the test with a 95\% significance level. %
NPSK does not require the assumptions of normal distributions, homogeneous distributions, and the minimum sample size~\cite{npsk-r}.

\section{Analysis}
\label{analyze-model}

\textbf{RQ1} motivates the need for incorporating visual code metrics into defect prediction models. While \textbf{RQ2} investigates if visual code metrics can enhance existing code prediction models, and \textbf{RQ3} explores if factors other than visual code metrics can affect the outcomes of defect prediction models.

\begin{figure*}[tbp]
\centering
\savebox{\largestimage}{\includegraphics[width=0.45\textwidth]{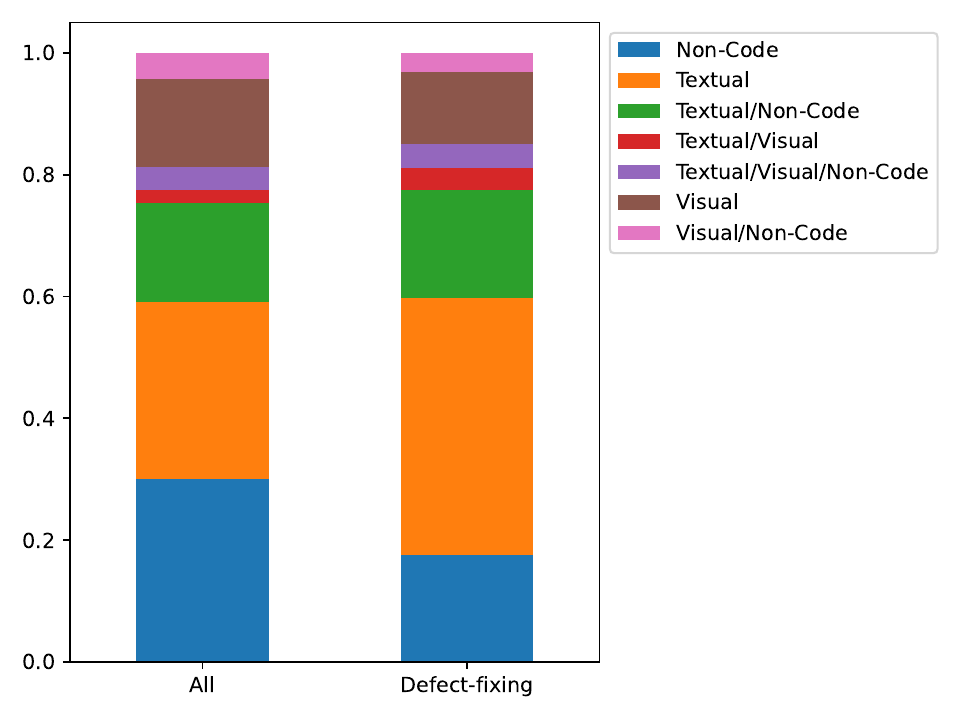}}%
\subcaptionbox[Short Subcaption]{%
    AAA video game project.%
    \label{subfig:ftc_1}%
}
[%
    0.45\textwidth %
]%
{\raisebox{\dimexpr.5\ht\largestimage-.5\height}%
{%
    \includegraphics[width=0.45\textwidth]{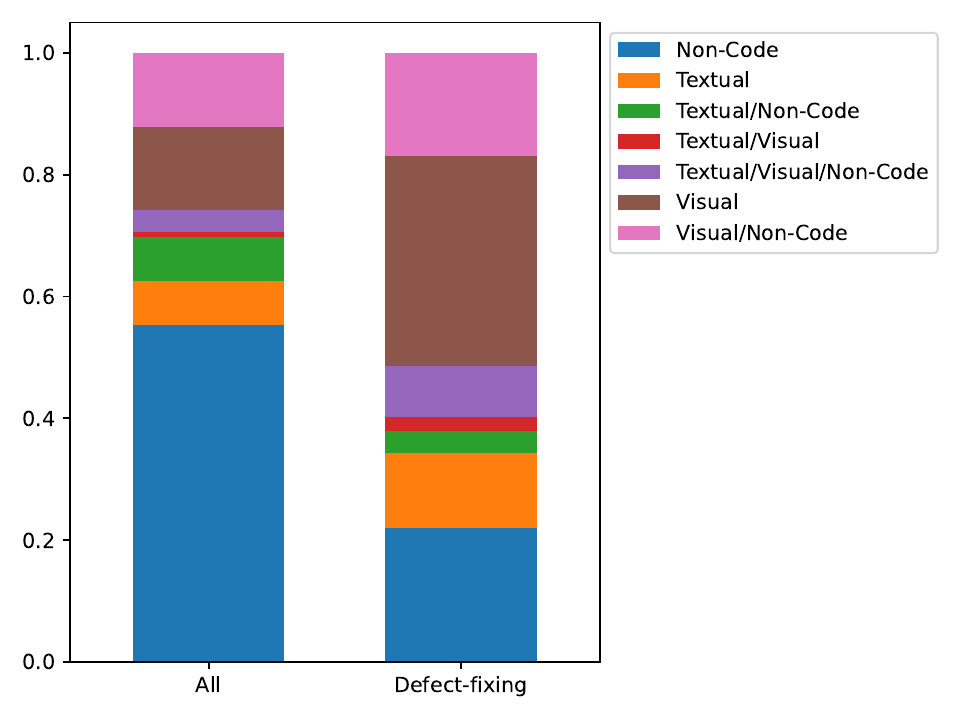}%
}}%
\hspace{0.05\linewidth} %
\subcaptionbox[Short Subcaption]{%
    Open source projects dataset.%
    \label{subfig:ftc_2}%
}
[%
    0.45\textwidth %
]%
{%
    \usebox{\largestimage}%
}%
\caption[Short Caption]{Percentage of commits containing different file types.}
\label{fig:file-types-commits}
\end{figure*}

\subsection*{{\normalfont\textbf{RQ1. Do defect-inducing commits with visual code occur more frequently than defect-inducing commits with textual code in the projects studied?}}}
We refer to \textit{textual code} as any code with a file extension that is able to be processed by the Python Lizard library~\cite{lizard} which contains a subset of the file extensions in Commit Guru~\cite{rosen2015commit}. We refer to \textit{visual code} as any code that can be parsed as visual code and contains file extension types related to the visual programming language at EA for the AAA video game and Max/MSP for the open source projects dataset.

There are 7 possible file type combinations in commits: (1) only non-code files, (2) only textual code, (3) only visual code, (4) textual code and non-code files, (5) visual code and non-code files, (6) textual code and visual code, and (7) textual code, visual code, and non-code files. These are of particular interest because the previous prediction model for the AAA video game only considered the textual combinations (1), (2), (4) meaning that important contextual information about visual code is missed. 

We present a stacked bar chart of the file type combination distribution for the 26,326 commits of the AAA video game project in \Cref{subfig:ftc_1}. For all commits, a majority contain only non-code files (55\%). However, with code files, there are more visual code files present (38\%) in commits than textual code files present (19\%) in commits. 

To understand what types of defects there are in the AAA video game project, we use the 5,296 defect-fixing commits as a proxy for the defects (i.e.\ a defect-fixing commit will change at least one defective file, hence it represents at least one defect by proxy). We visualize the breakdown of the 7 possible file type combinations of defect-fixing commits in \Cref{subfig:ftc_1}. We can see that there is a larger number of defect-fixing commits containing visual code (62\%) than defect-fixing commits containing textual code (27\%). Therefore, we can conclude that visual code defects occur more than textual code defect commits within this project. This motivates our investigation into incorporating visual code for defect prediction models.

For the open source projects dataset, we present an aggregated chart of the file type combinations in commits for the 64,246 commits of the 70 projects in \Cref{subfig:ftc_2}. Overall there are 30\% of commits that contain only non-code files, while 51\% of commits contain at least one textual code file and 25\% of commits contain at least one visual code file. We also look at the file types in defect-fixing commits and find that 67\% of commits contain textual code, while only 23\% of commits contain visual code. These ratios are different from those seen in the AAA video game project.

Seeing that the overall ratios of the open source projects dataset is different from the AAA video game project, we also see which of the 70 projects are individually similar to the breakdown of the AAA video game project where there are more commits with visual code files than textual code files. We find that 20 open source projects have more visual code files than textual files in all commits. While we find that 19 open source projects have more visual code files than textual files in defect-fixing commits.

\vspace{0.3cm}

\begin{tcolorbox}[colframe=gray!50]{\textit{Defects are occurring in visual code, so they need to be addressed.}}
\end{tcolorbox}

\vspace{0.1cm}

\subsection*{{\normalfont\textbf{RQ2. Do visual code features significantly improve the performance of defect prediction models for textual/visual code projects?}}}

This RQ is concerned with model performance and the effect of visual features on model performance.
To determine if visual code features can improve the performance of defect prediction models, we apply NPSK to the AUC and MCC results of each defect prediction model grouping by the Base, Textual, Visual, and Combined feature combinations described in \Cref{evalaute-model}.

\begin{figure*}[tb]
\centering
\savebox{\largestimage}{\includegraphics[width=0.45\textwidth]{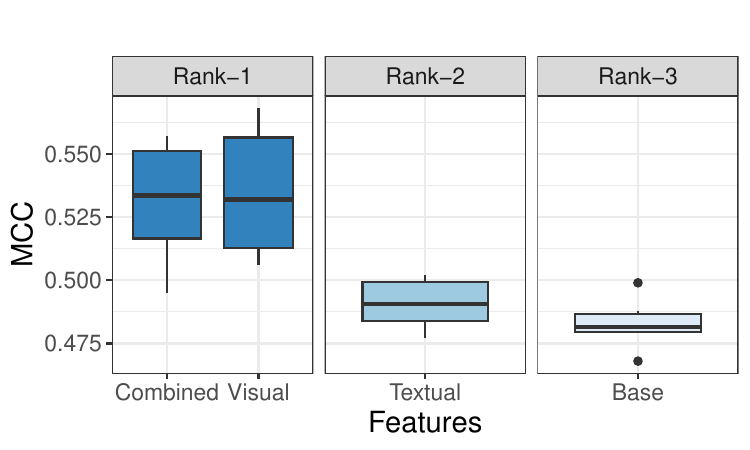}}%
\subcaptionbox[Short Subcaption]{%
    AUC%
    \label{subfig:i-auc_1}%
}
[%
    0.45\textwidth %
]%
{\raisebox{\dimexpr.5\ht\largestimage-.5\height}%
{%
    \includegraphics[width=0.45\textwidth]{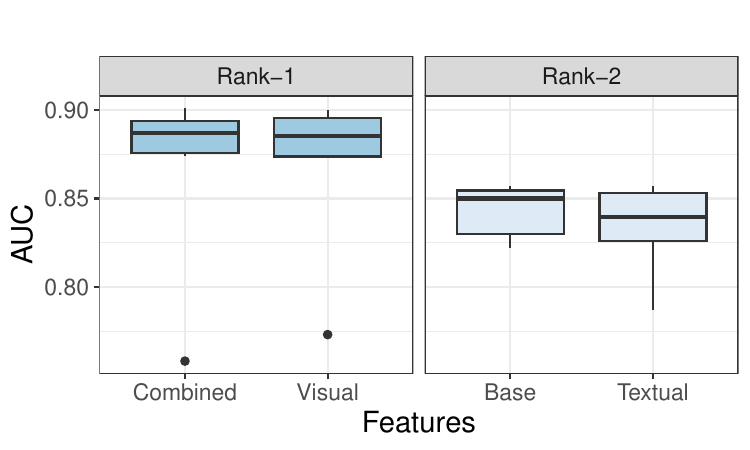}%
}}%
\hspace{0.05\linewidth} %
\subcaptionbox[Short Subcaption]{%
    MCC%
    \label{subfig:i-mcc_1}%
}
[%
    0.45\textwidth %
]%
{%
    \usebox{\largestimage}%
}%
\caption[Short Caption]{NPSK rankings of feature groups by AUC and MCC in AAA video game project.}
\label{fig:evaluation-auc-mcc-industry}
\end{figure*}

\begin{figure*}[tb]
\centering
\savebox{\largestimage}{\includegraphics[width=0.45\textwidth]{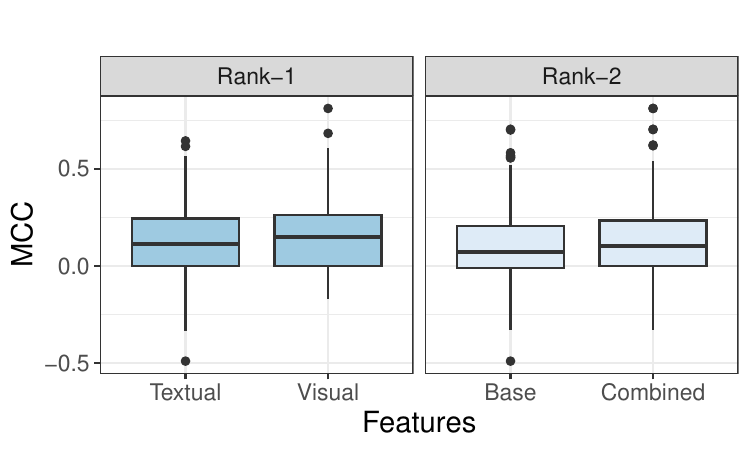}}%
\subcaptionbox[Short Subcaption]{%
    AUC%
    \label{subfig:auc_1-sf}%
}
[%
    0.45\textwidth %
]%
{\raisebox{\dimexpr.5\ht\largestimage-.5\height}%
{%
    \includegraphics[width=0.45\textwidth]{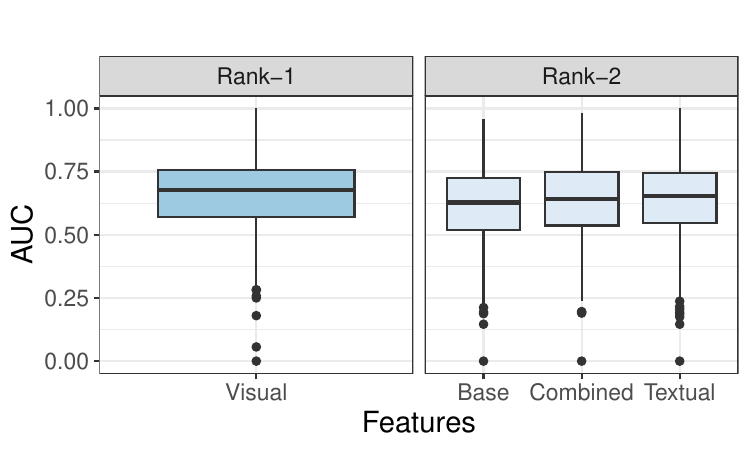}%
}}%
\hspace{0.05\linewidth} %
\subcaptionbox[Short Subcaption]{%
    MCC%
    \label{subfig:mcc_2-sf}%
}
[%
    0.45\textwidth %
]%
{%
    \usebox{\largestimage}%
}%
\caption[Short Caption]{NPSK rankings of feature groups by AUC and MCC in open source projects.}
\label{fig:evaluation-auc-mcc-opensource}
\end{figure*}
\begin{figure*}[tb]
\centering
\savebox{\largestimage}{\includegraphics[width=0.45\textwidth]{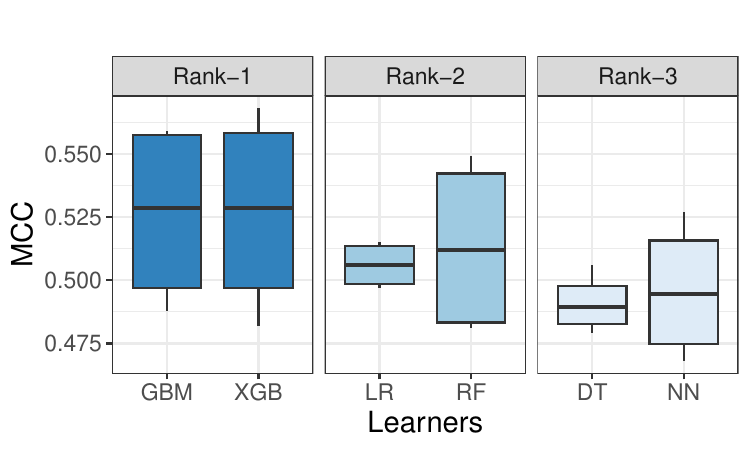}}%
\subcaptionbox[Short Subcaption]{%
    AUC%
    \label{subfig:mcc_1-learner-industry}%
}
[%
    0.45\textwidth %
]%
{\raisebox{\dimexpr.5\ht\largestimage-.5\height}%
{%
    \includegraphics[width=0.45\textwidth]{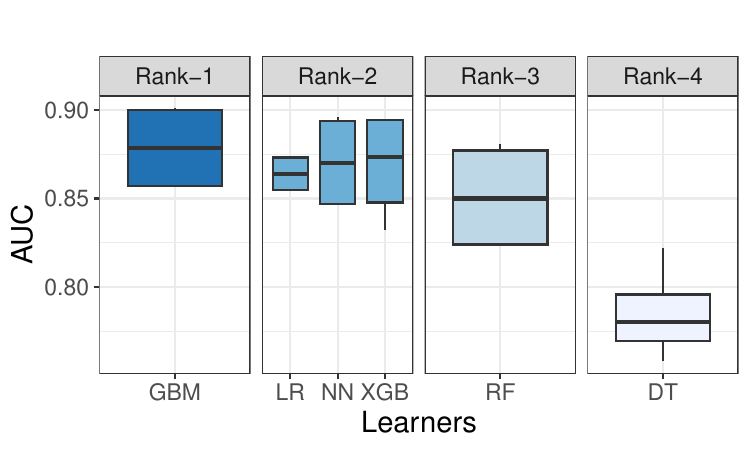}%
}}%
\hspace{0.05\linewidth} %
\subcaptionbox[Short Subcaption]{%
    MCC%
    \label{subfig:auc_2-learner-industry}%
}
[%
    0.45\textwidth %
]%
{%
    \usebox{\largestimage}%
}%
\caption[Short Caption]{NPSK rankings of learners by AUC and MCC in AAA video game project.}
\label{fig:learner-auc-mcc-industry}
\end{figure*}

\begin{figure*}[tb]
\centering
\savebox{\largestimage}{\includegraphics[width=0.45\textwidth]{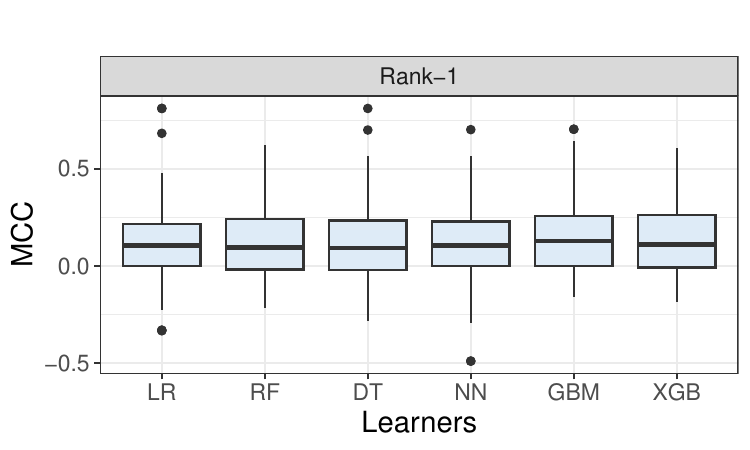}}%
\subcaptionbox[Short Subcaption]{%
    AUC%
    \label{subfig:mcc_1-open-learner}%
}
[%
    0.45\textwidth %
]%
{\raisebox{\dimexpr.5\ht\largestimage-.5\height}%
{%
    \includegraphics[width=0.45\textwidth]{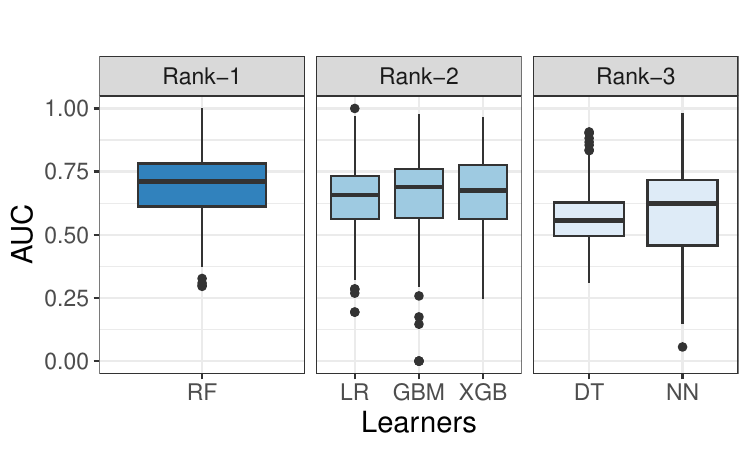}%
}}%
\hspace{0.05\linewidth} %
\subcaptionbox[Short Subcaption]{%
    MCC%
    \label{subfig:auc_2-learner-open}%
}
[%
    0.45\textwidth %
]%
{%
    \usebox{\largestimage}%
}%
\caption[Short Caption]{NPSK rankings of learners by AUC and MCC in open source projects.}
\label{fig:learner-auc-mcc-opensource}
\end{figure*}

\vspace{0.25cm}

We can see the results for the AAA video game project in \Cref{fig:evaluation-auc-mcc-industry}. In \Cref{subfig:i-auc_1}, the Combined and Visual feature combinations are ranked higher than the Base and Textual feature combinations for AUC meaning that visual code metrics contribute to a statistically significant difference in model prediction ability for the AAA video game project. We also can conclude that there is a significant difference in MCC, as we can see in \Cref{subfig:i-mcc_1} that the Combined and Visual feature combinations rank higher than the Base and Textual feature combinations.

\vspace{0.25cm}

The results of NPSK grouping by feature combinations for the open source projects can be seen in \Cref{fig:evaluation-auc-mcc-opensource}. NPSK is applied across projects and is used to determine if feature combinations are a significant factor in the group of projects. In terms of AUC, we can conclude that visual code feature combinations can help improve model performance as it ranks higher than the other feature combinations in \Cref{subfig:auc_1-sf}. In terms of MCC, we can conclude using \Cref{subfig:mcc_2-sf} that using only textual code feature combinations or using only visual code feature combinations can significantly improve model performance (MCC). However, when combining textual code and visual code features together, it will perform only as well as using no textual code and no visual features at all. 

Overall, we conclude that the addition of visual code features can improve AUC in both the AAA video game project and the open source projects. We also conclude that using visual code features can improve MCC for both the AAA video game project and the open source projects.

\begin{tcolorbox}[colframe=gray!50]{\textit{Contextual features for visual code improve defect prediction performance.}}
\end{tcolorbox}

\subsection*{{\normalfont\textbf{RQ3. Do the types of files in commits or choice of learners in projects significantly affect the outcome of defect prediction models?}}}

\vspace{0.3cm}

In RQ1, we find that the distribution of visual code files and textual code files in commits of projects can vary. 
Therefore, the first factor we consider is if projects have more commits containing visual code files or more commits containing textual code files. The second factor we consider is the 
choice of learners to train the defect prediction models for the Base, Textual, Visual, and Combined combination of features described in \Cref{evalaute-model}.

\vspace{0.3cm}

For the first factor, we split the projects into 2 groups where the \textit{first group} consists of 17 projects with more commits containing visual code files than textual code files and the \textit{second group} consists of 53 projects with more commits containing textual code files than visual code files. To determine if there is any significant difference among distributions of the AUC and MCC in the 2 groups, we perform the \textit{Wilcoxon rank-sum test} (WRST). The null hypothesis for the WRST test is that the distributions are not significantly different among each other. We reject the null hypothesis if $\text{p} < 0.05$. 

\vspace{0.2cm}

With the distributions of MCC in the 2 groups, we reject the null hypothesis ($\text{p} = 0.01 < 0.05$) meaning that there is a difference in a model's MCC performance among the 2 groups. With the distributions of AUC in the 2 groups, we do not reject the null hypothesis ($\text{p}=0.97 >0.05$) implying that the performance of models produced in each group do not significantly differ. However, there is no definitive conclusion about AUC being affected by majority visual code or majority textual code. Thus, we only conclude how well a model performs in terms of MCC can be affected by majority visual code or majority textual code.

For the second factor, we wish to see if the learners affect the outcomes of prediction models in terms of AUC and MCC. We apply NPSK to the MCC and AUC evaluation of each defect prediction model grouping by learner in \Cref{fig:learner-auc-mcc-industry} for the AAA video game project and in \Cref{fig:learner-auc-mcc-opensource} for the open source projects. 
We can see that the GBM (Gradient Boosting Method) learner model performs the best for both AUC and MCC in the AAA video game project. In the open source projects, we see that the RF (Random Forest) model performs the best for AUC, but the learner does not matter for MCC. From these results, we can conclude that learners do have a weak effect on the model prediction outcome, borne out only in AUC, not MCC. 

\begin{tcolorbox}[colframe=gray!50]{\textit{The types of files in commits, and learner choice matter for model performance, but perhaps not as much as contextual features do.}}
\end{tcolorbox}

\section{Threats to Validity}
Internal validity concerns the quality of labelled defective commits and bug reports used, how the features are chosen and extracted, and how the model is trained and tested which can affect the outcomes of our evaluation. To these extents, we may not have identified all potential defect-fixing commits, especially in our open source projects where we search commit messages for \textit{perfective} keywords. Furthermore, SZZ is imperfect~\cite{quach2021evaluating, fan2019impact} meaning that not all defect-inducing commits may have been identified with our implementations of SZZ.

Construct validity is if our conclusions follow from our assessments. We provide a rationale for our choice of performance measures and explore how file types in commits, and learners can affect model prediction outcomes with RQ2. We make careful use of statistical tools, such as NPSK, that address issues such as multiple hypothesis testing and non-parametric distributions.

External validity considers the extent to which the visual code defect prediction models can be applied to other projects that use visual programming languages, beyond those we selected for our evaluation. This validity is limited by the lack of open source projects with labelled defects that are similar to the AAA video game project. We attempt to address this by using more open source projects to explore the generality of our conclusions for the AAA video game project, but we are limited to just 2 visual programming languages. External validity is bolstered by the open release of our replication package for open source projects~\cite{reppackage}

\section{Conclusion}
In this paper, we present the use of visual code metrics for defect prediction to predict if a commit is defect-inducing. This work is motivated by the prevalent use of visual code to develop video game features at EA. We demonstrate how the performance of our current defect prediction models, which only consider process metrics and textual code metrics, could be improved by also including visual code metrics. We also conclude that visual code metrics improve the performance of defect prediction models for many open source projects. 

Future work includes investigating contextual visual code features such as dataflow, complexity measures, and code embeddings, to improve visual code defect prediction models.

The results of our research outline how incorporating visual code metrics into defect prediction models can help benefit JIT defect prediction at EA. To improve the replicability of our work, we release a replication package~\cite{reppackage} with 70 open source projects that contain visual code and textual code similar to the AAA video game project studied.

\bibliographystyle{IEEEtranN-ital}
\bibliography{main}

\end{document}